\documentclass[runningheads]{llncs}
\usepackage{graphicx}
\usepackage{epsfig}
\usepackage{todonotes}
\usepackage{hyperref}
\usepackage[ruled,vlined,linesnumbered,noresetcount]{algorithm2e}
\usepackage{framed}
\usepackage{amsmath}
\usepackage{float}
\newcommand{\spara}[1]{\smallskip\noindent{\bf{#1}}}

\begin{document}
\title{Distributed Computation of Top-$k$ Degrees ~\\ in Hidden Bipartite Graphs}

%
%


\author{Panagiotis Kostoglou\inst{1} \and
Apostolos N. Papadopoulos\inst{1} \and \\
Yannis Manolopoulos\inst{2}}

\authorrunning{P. Kostoglou et al.}
%

\institute{Aristotle University of Thessaloniki, Greece,\\
\email{panakost,papadopo@csd.auth.gr},
\and
Open University of Cyprus, Cyprus, \\
\email{yannis.manolopoulos@ouc.ac.cy}}

\maketitle              

\begin{abstract}
Hidden graphs are flexible abstractions that are composed of a set of known vertices (nodes), whereas the set of edges are not known in advance. To uncover the set of edges, multiple edge probing queries must be executed by evaluating a function $f(u,v)$ that returns either true or false, if nodes $u$ and $v$ are connected or not respectively. Evidently, the graph can be revealed completely if all possible $n(n-1)/2$ probes are executed for a graph containing $n$ nodes. However, the function $f()$ is usually computationally intensive and therefore executing all possible probing queries result in high execution costs. The target is to provide answers to useful queries by executing as few probing queries as possible. In this work, we study the problem of discovering the top-$k$ nodes of a hidden bipartite graph with the highest degrees, by using distributed algorithms. In particular, we use Apache Spark and provide experimental results showing that significant performance improvements are achieved in comparison to existing centralized approaches.
\end{abstract}

\section{Introduction}
\label{sec.intro}

Nowadays, graphs are frequently used to model real-life problems in many diverse application domains, such as social network analysis, searching and mining the Web, pattern mining in bioinformatics and neuroscience. Graph mining is an active and growing research area, aiming at knowledge discovery from graph data~\cite{Cook06,AW10}. In its simplest form, a graph $G(V,E)$ is defined by a set of nodes (or vertices) $V$ and a set of edges (or links) $E$. Each edge $e \in E$ connects a pair of nodes. In this work, we focus on \textit{simple bipartite graphs}, where the set of nodes is composed of two subsets $B$ (the \textit{black} nodes) and $W$ (the \textit{white} nodes), such as $V = B \cup W$ and $B \cap W = \emptyset$. Moreover, all edges of $G$ connect nodes from $B$ and nodes of $W$, meaning that the two endpoints of an edge belong to different node subsets. 

Bipartite graphs have many interesting applications in diverse fields. For example, a bipartite graph may be used to represent product purchases by customers. In this case, an edge exists between a product $p$ and a customer $c$ when $p$ was purchased by $c$. As another example, in an Information Retrieval or Text Mining application a bipartite graph may be used to associate different types of tokens that exist in a document. Thus, an edge between a document $d$ and a token $t$ represents the fact that the token $t$ appears in document $d$.
Moreover, there are many applications of bipartite graphs in systems biology and medicine~\cite{PKP+14}.

\begin{figure}[!ht]
\begin{center}
\includegraphics[scale=0.5]{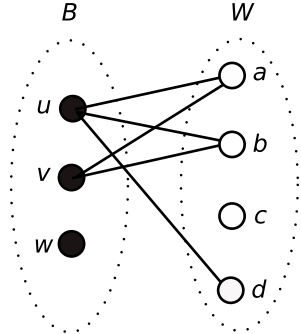}
\end{center}
\caption{An example of an undirected bipartite graph, with two bipartitions $B=\{u,v,w\}$ and $W=\{a,b,c,d\}$.}
\label{fig.bipartite}
\end{figure}

Figure~\ref{fig.bipartite} depicts a simple bipartite graph containing seven nodes and five edges. Among the black nodes, $u$ has the maximum number of neighbors (i.e., three). One of the nodes is isolated, i.e., it does not have any neighbors. 
~\\

\spara{Motivation and Contributions}.
Conventional bipartite graphs are characterized by the fact that the sets of vertices $B$ and $W$ as well as the set of edges $E$ are known in advance. Nodes and edges are organized in a way to enable efficient execution of fundamental graph-oriented computational tasks, such as finding nodes with large degree, computing shortest paths, clustering and community detection. Usually, the adjacency list representation is being used, which is a good compromise between space requirements and computational efficiency. However, a concept that has started to gain significant interest recently is that of \textit{hidden graphs}. In contrast to a conventional graph, a hidden bipartite graph is defined as $G(B,W,f())$, where $B$ and $W$ are the subsets of nodes and $f()$ is a function $B \times W$ $\rightarrow$ $\{0,1\}$ which takes as an input two vertex identifiers and returns true or false if the edge exists or not respectively. Therefore, in a hidden graph the edge set $E$ is not given explicitly and it is inferred by using the function $f()$. 

Hidden graphs are very powerful because they are able to represent any type of relationship among a set of entities. Formally, for $n$ nodes, there is an exponential number of different graphs that may be generated (the exact number is $2^{\binom{n}{2}}$). To materialize all these different graphs demands significant space requirements. Moreover, it is unlikely that all these graphs will be used eventually. In contrast, the use of a hidden graph enables the representation of any possible graph by the specification of the appropriate function $f()$. It is noted that the function $f()$ may require the execution of a complex algorithm in order to decide if two nodes are connected by an edge. Therefore, interesting what-if scenarios may be tested with respect to the satisfaction of meaningful properties.  
It is evident, that the complete graph structure may be revealed if all possible $n_b \times n_w$ edge probing queries are executed, where $n_b = |B|$ and $n_w = |W|$. However, this solution involves the execution of a quadratic number of probes, which is expected to be highly inefficient, taking into account that the function $f()$ is computationally intensive and that real-world graphs are usually very large. Therefore, the target is to provide a solution to a graph-oriented problem by using as few edge probing queries as possible. 
The problem we are targeting is the discovery of the $k$ nodes with the highest degrees. The degree of a node $v$ is defined as the number of neighbors of $v$. This problem has been addressed previously by~\cite{TSL10} and \cite{YLW13}. 
In this work, we are interested in solving the problem by using distributed algorithms and over Big Data architectures. This enables the analysis of massive hidden graphs and will provide some baseline for executing more complex graph mining tasks, thus overcoming the limitations provided by centralized approaches. In particular, we study distributed algorithms for the discovery of the top-$k$ degree nodes in Apache Spark~\cite{} over YARN and HDFS~\cite{White15} and we offer experimental results based on a cluster of 32 physical machines. Performance evaluation results demonstrate that the proposed techniques are scalable and can be used for the analysis of massive hidden networks. We note that this is the first work addressing the problem of hidden graph analysis in a distributed setting.   
~\\

\spara{Roadmap}. The rest of the paper is organized as follows. The next section contains the most representative related work in the area. Section~\ref{sec.proposed} studies our main contribution in detail. Performance evaluation results are offered in Section~\ref{sec.performance}, whereas Section~\ref{sec.conclusions} concludes our work and presents briefly interesting directions for future work in the area.

\section{Related Research}
\label{sec.related}

A hidden graph is able to represent arbitrary relationship types among graph nodes. A research direction that is strongly related to hidden graphs is \textit{graph property testing}~\cite{GGR98}. In this case, one is interested in detecting if a graph satisfies a property or not by using fast approximate algorithms. 

Evidently, detecting graph properties efficiently is extremely important. However, an ever more useful and more challenging task is to detect specific subgraphs or nodes that satisfy specific properties, by using as few edge probing queries as possible. 

Another research direction related to hidden graphs focuses on \textit{learning} a graph or a subgraph by using edge probing queries using pairs or sets of nodes (group testing) \cite{AV05}. A similar topic is the \textit{reconstruction} of subgraphs that satisfy certain structural properties \cite{BGK05}. 

An important difference between graph property testing and analyzing hidden graphs is that in the first case the graph is known whereas in the second case the set of edges is unknown and must be revealed gradually. Moreover, property testing focuses on checking if a specific property holds or not, whereas in hidden graph analysis we are interested in answering specific queries exactly. The problem we attack is the discovery of the top-$k$ nodes with the highest degrees. This problem was solved by~\cite{TSL10} and~\cite{YLW13} assuming a centralized environment. 

The basic algorithm used in~\cite{TSL10} was extended in~\cite{Strouthopoulos2017} for unipartite graphs, towards detecting if a hidden graph contains a $k$-core or not. Again, the algorithm proposed in~\cite{Strouthopoulos2017} is centralized and its main objective is to minimize the number of edge probing queries performed. 

In this paper, we take one step forward towards the design and performance evaluation of distributed algorithms for detecting the top-$k$ nodes with the highest degrees in hidden bipartite graphs, aiming at the reduction of the overall computational cost which depends on both the number of edge probing queries and the level of parallelism used. We note that this is the first work in the literature to attack the specific problem in a distributed setting.

\section{Fundamental Concepts and Problem Definition}
\label{sec.fundamentals}

In this section, we present some fundamental concepts related to our research and state the problem formally. Recall that, the detection of the top-$k$ nodes with the highest degrees in a bipartite graph has been attached in~\cite{TSL10} taking a centralized perspective. In that paper, the authors present the \textit{Switch-On-Empty} (SOE) algorithm which shows the best performance and provide an optimal solution with respect to the number of edge probing queries that are required. Table~\ref{tab.symbols} presents the most frequently used symbols.

\begin{table}[!h]
\begin{center}
\caption{Frequently used symbols.}
\label{tab.symbols}
\renewcommand{\arraystretch}{1.2}
\begin{tabular}{|c||l|}
\hline
{\bf Symbol}    & {\bf Interpretation}                   				        	\\ \hline\hline
$G(B,W,f)$		& a hidden graph											        \\ \hline
$B$				& set of black vertices of $G$										\\ \hline
$W$				& set of black vertices of $G$										\\ \hline
$f(b,w)$	    & {\it true} if the edge $(b,w)$ exists, {\it false} otherwise 	    \\ \hline
$n_b$, $n_w$	& number of black and white vertices 								\\ \hline
$k$				& number of highest degree vertices requested 						\\ \hline
$N(u)$			& set of neighbors of vertex $u$									\\ \hline
$d(u)$			& degree of vertex $u$  											\\ \hline
$E$				& (unknown) set of edges 											\\ \hline
$m$				& (unknown) number of edges ($m=|E|$)								\\ \hline
$s(u)$			& number of known existing neighbors of $u$							\\ \hline
$e(u)$			& number of known non-existing neighbors of $u$						\\ \hline
$probes$		& total number of edge probing queries issued						\\ \hline
\end{tabular}
\end{center}
\end{table}

Before diving into the details of our proposal, there is a need to describe briefly the way the SOE algorithm works.
SOE receives as input a hidden bipartite graph, with bipartitions $B$ and $W$. The output of SOE is composed of the $k$ vertices from $B$ or $W$ with the highest degrees. Without loss of generality, we are focusing on vertices of $B$. Edge probing queries are executed as follows: 

\begin{itemize}
\item
Initially, SOE starts from a vertex $b_1 \in B$, selects a vertex $w_1 \in W$ and executes $f(b_1, w_1)$. If the edge $(b_1,w_1)$ is solid, it continues to perform probes between $b_1$ and another vertex $w_2 \in W$. 
\item
Upon a failure, i.e., when the probe $f(b_1, w_j)$ returns an empty result, the algorithm applies the same process for another vertex $b_2 \in B$. Vertices for which all the probes have been applied, do not participate in future edge probes. 
\item
A round is complete when all vertices of $B$ have been considered. After each round,
some vertices can be safely included in the result set $R$ and they are removed from $B$. 
When a vertex $b_i$ must be considered again, we continue the execution of probes remembering the location of the last failure. 
\item
SOE keeps on performing rounds until the upper bound of vertex degrees in $B$ is less than the current $k$-th highest degree determined so far. In that case, $R$ contains the required answer and the algorithm terminates. Note that all equal-degree vertices are included in $R$.
\end{itemize}

In this work, we proceed one step further in order to detect high-degree nodes in larger graphs using multiple resources. In this respect, our algorithms are implemented in the Apache Spark engine~\cite{Zaharia12,Zaharia16} using the Scala programming language~\cite{Odersky16}. Spark offers a powerful environment that enables the execution of distributed applications in massive amounts of data.  
Apache Spark is a unified distributed engine with a rich and powerful API for Scala, Python, Java and R~\cite{Zaharia16}. One of its main characteristics is that (in contrast to Hadoop MapReduce) it exploits main memory as much as possible, being able to persist data across rounds to avoid unnecessary I/O operations. Spark jobs are executed based on a master-slave model in cooperation with a cluster manager such as YARN~\footnote{http://hadoop.apache.org/} or MESOS~\footnote{http://mesos.apache.org/}.


\section{Proposed Approach}
\label{sec.proposed}
In the following sections, we will focus on the algorithms developed to solve the distributed detection of top-$k$ nodes in a hidden bipartite graph. Two algorithms are presented that are both inspired by the SOE (Switch-On-Empty) algorithm~\cite{TSL10}. The first algorithm, Distributed Switch-On-Empty (DSOE), is a distributed version of the SOE algorithm. The second algorithm, , DSOE$^*$, is an improved and more flexible version of DSOE. 

\subsection {The DSOE Algorithm}
Our first solution is very simple but at the same time very effective. The main mechanism has the following rationale. For a vertex $b \in B$ we are executing edge-probing queries until we get a certain amount of negative results $f$. The value of $f$ is relevant to the size of the graph and is being reduced exponentially. More precisely, the first batch will have $f = |W|/2$ the second $f = |W|/4$ and so on. The idea behind this is that, because the majority of the degree distribution in real life graphs follows a Power law distribution, we do not expect to find many vertices with a high degree. The closer we get to the point of exhausting all $w \in W$ the smaller we set $f$, because we want to avoid as many unnecessary queries as possible. The pseudocode of the routine is given below. 
~\\

\SetAlgoLined
\SetKwProg{Fn}{Function}{ is}{end}
\Fn{routine(v: vertex, f: Int) : void}{
 \While{$(False probes) \leq f$}{
  Execute the next query \;
  
  \uIf{edge-probing query==False}{
    $False probes++$\;}

 }}
~\\

The routine can be implemented easily to work in a distributed setting of the form $\forall b \in B$. After each batch of routines, we check for the vertices that belong to the answer set $R$. Vertices that belong to $R$ can be recognized easily, since vertices that have completed all possible probes $\forall w \in W$ should be in the answer. If $|R| \geq k$, then we need one last batch of routines to finalize $R$. 

\begin{algorithm}[H]
\SetAlgoLined
\KwResult{Return a set $R$ of vertices that have the highest degree }
 $f=2$\;
 \While{$\mid R \mid \leq k$}{
  map(routine($\forall b \in B$,$f$)) \;
  $\forall b \in B$ we check if b has probe all the vertices $\forall w \in W$,each vertices that did it is being removed from $B$ and is being added to $R$\;
  $f=f*2$
 }
 map(routine($\forall b \in B$,$f$))\;
  $\forall b \in B$ we check if b has probed all the vertices $\forall w \in W$,each vertices that did it, is being removed from $B$ and is being added to $R$\;
 return R\;
 \caption{DSOE}
\end{algorithm}

In this part we will prove that our conditions for the loop and for the vertices that belong in $R$ are correct. For a vertex $b_1 \in B$, if $b_1$ completes all possible probes from the first routine, then there are only two options for it's degree: either $d(b_1) = |W|$ or $d(b_1) = |W| - 1$, because there is a possibility that the last query could be negative and still $b_1$ could have exhausted all possible probes. Assume that while we execute the loop we are in the $t_1$ repeat. Also we assume that after $t_1$ we get a vertex $b_2 \in B$ that has exhausted all possible edge-probing queries. For $d(b_2)$ again we have two possibilities, either $d(b_2) = |W| - t_1$ or $d(b_2) = |W| - t_1 -1$ for the same reason as previously. 

From the previous observations we can safely conclude that for a vertex $b3 \in B$, that has completed all possible queries in $t_2$ with $t_2 > t_1$ the best-case scenario for the degree is $d(b_3) = |W| - t_2$ and the worst case scenario is $d(b_2) = |W| - t2 -1$. 

It is evident that $d(b_2) \geq d(b_3)$ and the equality is possible only if $t_2 = t_1 + 1$. From this conclusion we are sure that if $|R| \leq k$ then if a vertex $b \in B$ exhausts all the queries, it can be inserted into $R$. Taking into concern the equality we have to run the batch of routines one last time for the possibility that we may find a vertex with degree equal to the minimum degree in $R$. Algorithm 1 contains the outline of DSOE.

\subsection{The DSOE$^*$ Algorithm}
DSOE is the natural extension to the centralized SOE algorithm. However, we advance one step further in order to improve runtime as much as possible, even if the number of probes increases. For very large graphs we already expect a very large number of probes that is inevitable. For that reason trying to improve the execution time by adding a very small amount of probes in the already big number of inevitable probes may worth for the overall performance of our algorithm. The updated algorithm is named DSOE$^*$.

DSOE treate all vertices equally, so for algorithms we want to make an initial prediction about the degree of the vertices and after that, to emphasize during the routines the vertices that we predict to have a large degree by having a more loose condition in contrast to those that we predict that will have small degree. The result of this process is to calculate all possible edge-probing queries for the vertices we predict to have a large degree. With the information of the exact number of the degree for those vertices we are able to have a better ending condition for the rest of the vertices. Therefore, we will be able to stop processing of the vertices with low degrees faster and this way we will have better execution times. Moreover, for real life graphs, the degrees of the vertices usually follow a Power law distribution,so we expect that most vertices of the graph will have way smaller degree in comparison with the degree of the top-$k$ vertices. 

More specifically, DSOE$^*$ initially performs some random edge-probing queries $\forall b \in B$. The number for these queries is set to $\log(\log(|W|))$. After this step, the course of DSOE$^*$ is quite similar to that of DSOE. We execute repeatedly a batch of routines $\forall b \in B$, with the difference that, this time a single routine for a vertex $b \in B$ is complete when $N$ negative edge-probing queries, where $N$ is an outcome of the prediction performed by the sampling process to quickly detect nodes that potentially have large degree.

When a vertex exhausts all possible queries, then it is added gets added to a  temporary set $M$. The vertices in the set $M$ may not be completely correct, so we have to continue with the probing queries. However, the contents of $M$ provide a threshold $T$ which is set to the minimum degree among the vertices in $M$. Probing queries are executed $\forall b \in B$, until $d(b)<T$ . The vertices that have completed all possible probes are added to $M$. Finally, the $k$ best vertices from $M$ with respect to the degree are returned as the final answer. Algorithm 2 contains the pseudocode of DSOE$^*$. ~\\

\SetAlgoLined
\SetKwProg{Fn}{Function}{ is}{end}
\Fn{prediction(v: vertex,probes: int) : int}{
\While{counter $<$ probes}{
\uIf{edge-probing query==True}{
prediction$++$\;}}
return prediction\;
}

\SetAlgoLined
\SetKwProg{Fn}{Function}{ is}{end}
\Fn{routine(v: vertex) : void}{
\While{$negatives < v.prediction$}{
\uIf{edge-probing query==False}{
negatives$++$\;}}
}

\SetAlgoLined
\SetKwProg{Fn}{Function}{ is}{end}
\Fn{exhaust(v: vertex,T: int) : void}{
\While{$T \leq v.degree$}{
execute the next edge-probing query\;
}}
~\\

\begin{algorithm}[H]
 \caption{DSOE$^*$}
 \KwResult{Return a set $R$ of vertices that have the highest degree }
 map(prediction($\forall b \in B$,$\log(\log(\mid W \mid))$))\;
 \While{$\mid M \mid \leq k$}{
  map(routine($\forall b \in B$))\;
  $\forall b \in B$ we check if b has probe all the vertices $\forall w \in W$,each vertices that did it is being removed from $B$ and is being added to $M$\;
 }
 threshold=min(M)\;
 map(exhaust($\forall b \in B$,threshold))\;
 $\forall b \in B$ we check if b has probe all the vertices $\forall w \in W$,each vertices that did it is being removed from $B$ and is being added to $M$\;
 $R$=top-$k$ from $M$\;
 return $R$\;
\end{algorithm}

\section{Performance Evaluation}
\label{sec.performance}

In this section, we present performance evaluation results depicting the efficiency of the proposed distributed algorithms. All experiments have been conducted in a cluster of 32 physical nodes (machines) running Hadoop 2.7 and Spark 2.1.0. One node is used as the \textit{master} whereas the rest 31 nodes are used as \textit{workers}. The data resides in HDFS and YARN is being used as the resource manager.

\subsection{Datasets}
All datasets used in performance evaluation correspond to real-world networks. 
The networks used have different number of black and white nodes as well as different number of edges. More specifically, we have used three networks: DBLP, YOUTUBE, and WIKI. All networks are publicly available at the Koblenz Network Collection, which is a project to collect large network datasets of different types, in order to assist research in network science and related fields. The collection is maintained by the Institute of Web Science and Technologies at the University of Koblenz–Landau and it accessible by the following URL: \url{http://konect.uni-koblenz.de/}.
\begin{itemize}
    \item 
    \textbf{DBLP}. The DBLP network is the authorship network from the DBLP computer science bibliography. The network is bipartite and a node is either an author or a publication.  Each edge connects an author to one of the publications. The characteristics of this network are as follows: $|B| = 4,000,150$, $|W| = 1,425,813$, the highest degree for set $B$ is 114, the highest degree for set $W$ is 951 and the average degree for $B$ is 6.0660 and for $W$ is 2.1622. 
    \item
    \textbf{YOUTUBE1}. This is the bipartite network of YouTube users and their group memberships. Nodes correspond to users and groups, and an edge between a user and a group denotes a group membership. The properties of this network are as follows: $|B| = 94,238$ and $|W| = 30,087$, the highest degree of set $B$ is $1,035$, the highest degree of set $W$ is $7,591$, the average degree for $B$ is $3.1130$ and the average degree of $W$ is $9.7504$.
    \item
    \textbf{YOUTUBE2}. This network represents friendship relationships between YouTube users. Nodes are users and an undirected edge between two nodes indicates a friendship. Although this graph is unipartite we modified it in order to become bipartite. We duplicate it's vertices and the set $B$ is one clone of the nodes whereas $W$ is the other. Two nodes between $B$ and $W$ are connected if their corresponding nodes are connected in the initial graph. Evidently, for this graph the statistics for the two sets $B$ and $W$ are exactly the same: $|B| = |W| = 1,134,890$, the maximum degree is $28,754$ for both sets and the average degree for both sets is $5.2650$ 
    \item
    \textbf{WIKIPEDIA}. This is the bipartite network of English Wikipedia articles and the corresponding categories they are contained in. The first set of nodes corresponds to articles and the second corresponds to categories. For this graph we have $|B|= 182,947$, $|W|= 1,853,493$, the highest degree for $B$ is $11,593$ and for $W$ is $54$, the average degree for $B$ is $2.048$ while for set $W$ is $20.748$.
\end{itemize}

\subsection{Experimental Results}
\begin{figure*}[!h]
\begin{center}
\begin{minipage}{6cm}
\centerline{\epsfig{file=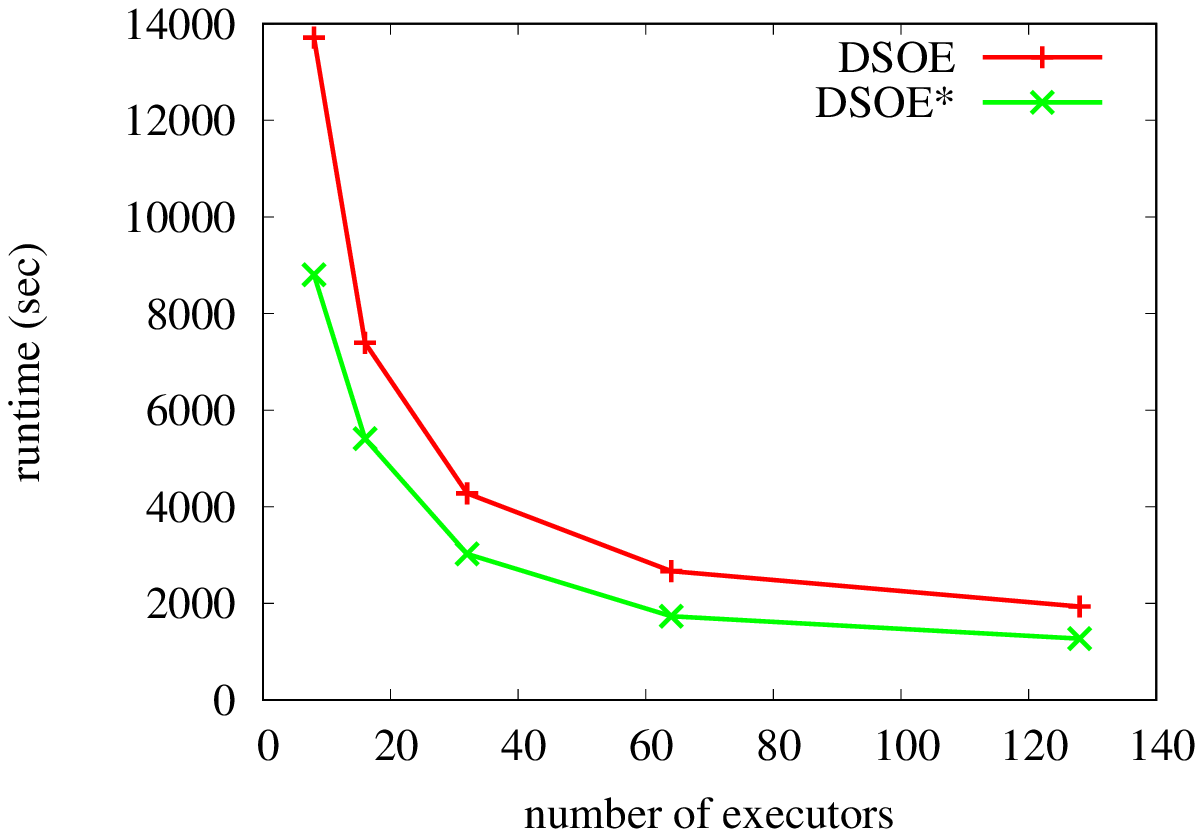,width=6cm}}
\centerline{(a) runtime vs number of executors}
\end{minipage}
\begin{minipage}{6cm}
\centerline{\epsfig{file=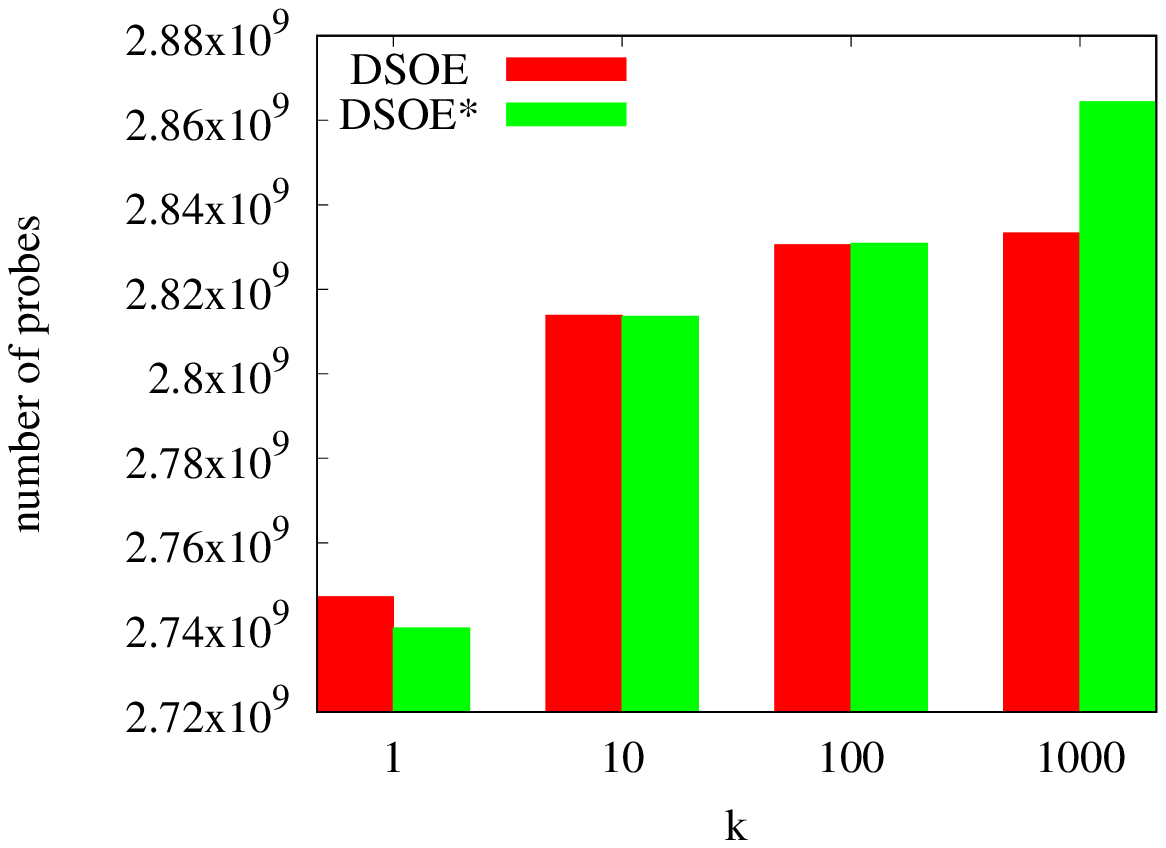,width=6cm}} 
\centerline{(b) number of probes vs $k$}
\end{minipage}
\end{center}
\caption{Comparison of DSOE and DSOE$^*$ with respect to runtime and number of probes.}
\label{fig.timeandprobes}
\end{figure*}

In the sequel, we present some representative experimental results demonstrating the performance of the proposed techniques. First, we perform a comparison of DSOE and DSOE$^*$ with respect to runtime (i.e., time duration to complete the computation in the cluster) by modifying the number of Spark executors running. For this experiment, we start with 8 Spark executors  
and gradually we keep on increasing their number keeping $k=10$. The DBLP graph has been used in this case. Figure \ref{fig.timeandprobes}(a) shows the runtime of both algorithms by increasing the number of executors. It is evident that both algorithms are scalable, since there is a significant speedup as we double the number of executors. Moreover, DSOE$^*$ shows better performance than DSOE. 

Another important measure that we want to examine is the number of probes that both algorithms execute. For this experiment, the number of executors and the execution time are irrelevant, so we focus only on different values of $k$. Moreover we are running our algorithms in a significantly smaller dataset (YOUTUBE1) for different values of $k$ (i.e., 1, 10, 100 and 1000). The corresponding results are given in Figure~\ref{fig.timeandprobes}(b). The first observation is that both algorithms perform a significant amount of probes. However, this was expected taking into account that if the average node degree is very small or very large, then many probes are required before the algorithms can provide the answer, as it has been shown in~\cite{TSL10}. Also, in general DSOE$^*$ requires less probes in comparison to DSOE for small values of $k$. 

\begin{figure*}[!h]
\begin{center}
\begin{minipage}{6cm}
\centerline{\epsfig{file=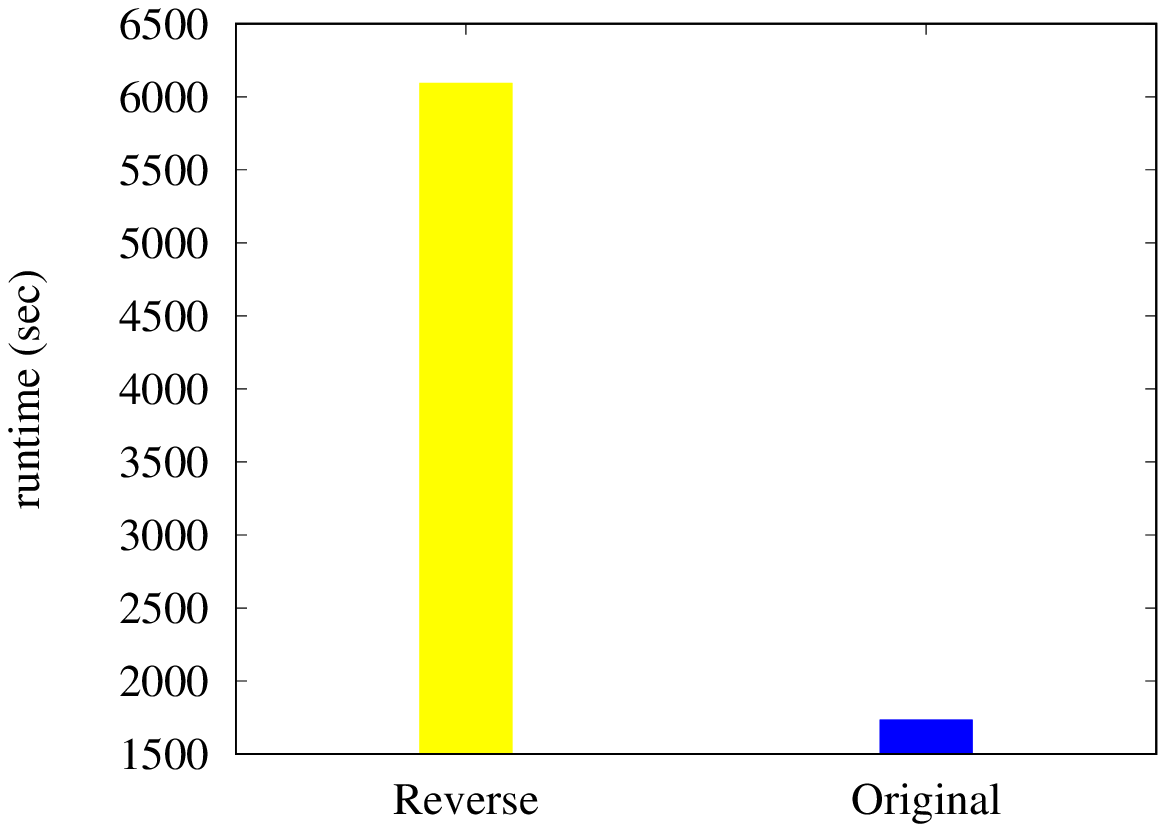,width=6cm}}
\centerline{(a) runtime vs source set}
\end{minipage}
\begin{minipage}{6cm}
\centerline{\epsfig{file=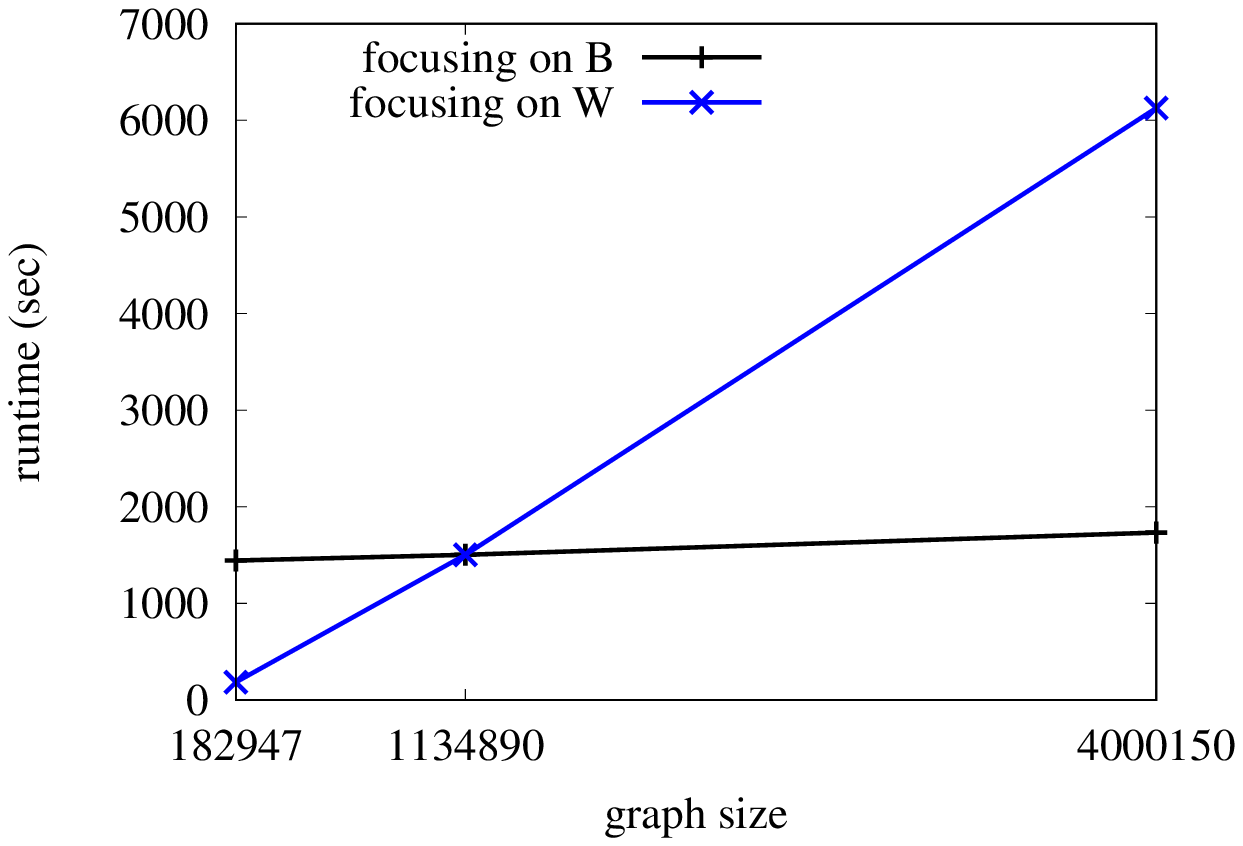,width=6cm}} 
\centerline{(b) runtime vs cardinality of $B$ and $W$}
\end{minipage}
\end{center}
\caption{Performance of DSOE$^*$ with respect to the source nodes and the size of the node sets $B$ and $W$.} 
\label{fig.results2}
\end{figure*}

One more very interesting aspect that we examine is the performance of our approaches with respect to the size of the graph. Our motivation is to compare the impact of sets $B$ and $W$ to the execution time. For this reason we will use DSOE$*$ on the DBLP graph twice: for the first execution we have $|B| = 4,000,150$ and $|W| = 1,425,813$ and the for the second time we reverse the direction of the queries from $W$ to $B$. This way the graphs that we compare have the exact same number of edges and vertices but the differ significantly in the cardinality of $B$ and $W$. Figure~\ref{fig.results2}(a) presents the corresponding results. It is observed that in the reverse case the runtime is significantly higher, since for every node in $W$ there are more options to perform probes on $B$. In general, the cost of the algorithm drops if the cardinality if $B$ is larger than that of $W$.  

The goal of the last experiment is to test the scalability of the algorithms by increasing the size of the data. First we focus on $B$ and then on $W$. For this experiment DBLP, YOUTUBE2 and WIKIPEDIA datasets have been used.  These graphs have almost the same cardinality in one set and they differ on the cardinality of the other. For all tests we have used 64 Spark executors and the implementation of the DSOE$*$ algorithm with $k=10$. The corresponding results are given in Figure~\ref{fig.results2}(b). It is evident that although the cardinalities of both $B$ and $W$ have an impact on peformance, execution time is more sensitive on the cardinality of $W$ when used as the source set.





\section{Conclusions}
\label{sec.conclusions}
In this work, we study for the first time the distributed detection of the top-$k$ nodes with the highest degrees in a hidden bipartite graph. Since the set of edges is not available a-priori, edge probing queries must be applied in order to be able to explore the graph. We have designed two algorithms to attack the problem (DSOE and DSOE$^*$) and evaluate their performance based on real-world networks. In general the algorithms are scalable, showing good speedup factors by increasing the number of Spark executors. 

There are many interesting directions for future work. By studying the experimental results, one important observation is that the number of probes is in general large. Therefore, more research is required to be able to significantly reduce the number of probes preserving the good quality of the answer. Also, it is interesting to evaluate the performance of the algorithms in even larger networks containing significantly larger number of nodes and edges.



\section*{Acknowledgments}
The authors would like to thank the DaSciM group of the LiX laboratory of Ecole Polytechnique, and specifically Prof. Michalis Vazirgiannis and Dr. Christos Giatsidis for sharing the cluster to conduct the experimental evaluation.

%
\bibliographystyle{splncs04}
\bibliography{mybib}

\end{document}